\input epsf
\documentstyle[prb,aps]{revtex}
\begin{document}
\draft
\title{QUASI-LONG-RANGE ORDER IN RANDOM-ANISOTROPY HEISENBERG MODELS}
\author{Ronald Fisch}
\address{Department of Physics\\
Washington University\\
St. Louis, Missouri 63130}
\date{Submitted to Physical Review B, 5 January 1998}
\maketitle
\begin{abstract}
Monte Carlo simulations have been used to study a discretized Heisenberg
ferromagnet (FM) with random uniaxial single-site anisotropy on
$L \times L \times L$ simple cubic lattices, for $L$ up to 64.  The spin
variable on each site is chosen from the twelve [110] directions.  The
random anisotropy has infinite strength and a random direction on a fraction
$x$ of the sites of the lattice, and is zero on the remaining sites.  In
many respects the behavior of this model is qualitatively similar to that
of the corresponding random-field model.  Due to the discretization, for
small $x$ at low temperature there is a [110] FM phase.  For $x>0$ there is
an intermediate quasi-long-range ordered (QLRO) phase between the paramagnet
and the ferromagnet, which is characterized by a $|{\bf k}|^{-3}$ divergence
of the magnetic structure factor S$({\bf k})$ for small $\bf k$, but no true
FM order.  At the transition between the paramagnetic and QLRO phases
S$({\bf k})$ diverges like $|{\bf k}|^{-2}$.  The limit of stability of the
QLRO phase is somewhat greater than $x=0.5$.  For $x$ close to 1 the low
temperature form of S$({\bf k})$ can be fit by a Lorentzian, with a
correlation length estimated to be $11 \pm 1$ at $x=1.0$ and $25 \pm 5$ at
$x=0.75$.
\end{abstract}
\section{INTRODUCTION}

The Heisenberg model with random uniaxial single-site anisotropy is
considered to be the proper model for studying amorphous
alloys\cite{PRA,HAS} of non-$S$-state rare earths (RE) and transition metals
(TM), such as Tb$_x$Fe$_{1-x}$.  The model was introduced by Harris,
Plischke and Zuckermann,\cite{HPZ} who performed a mean-field calculation
and found a ferromagnetic (FM) phase at low temperature.  It was shown later
by Pelcovits, Pytte and Rudnick,\cite{PPR} using an argument parallel to
that of Imry and Ma\cite{IM} for the random field case, that such a FM phase
is not stable in three dimensions.

The actual behavior of this model in three dimensions has remained a subject
of controversy.  It was argued by some workers\cite{PPR,CL,CSS,Cha} that
there should be a low temperature Ising spin-glass phase, but the numerical
evidence for this was never convincing.\cite{JK,Fi1}  It has recently been
shown by Migliorini and Berker\cite{MB} that in three dimensions the Ising
spin glass is destabilized by a random field.  The spin glass phase, as it
is usually envisioned, has spontaneously broken time-reversal symmetry
(i.e. the time-average expectation values of local moments do not vanish).
Therefore, one would expect that it should also be destabilized by random
uniaxial anisotropy.

An alternative scenario, first proposed by Aharony and Pytte,\cite{AP} is
that at low temperature there is a phase characterized by an infinite
magnetic susceptibility and a power-law decay of two-spin correlations as a
function of the distance between the two spins, but no true FM long-range
order.  This power-law decay of the two-spin correlations is called
quasi-long-range order (QLRO).  In the approximation of Aharony and Pytte,
and its later elaboration by Goldschmidt and Aharony,\cite{GA} it is claimed
that the infinite-susceptibility phase occurs in the presence of random
anisotropy, but not in the presence of a random field.  This is rather
problematic, as it requires that the infinite-susceptibility phase not break
time-reversal symmetry.  The relationship between the ordered states of
random-field and random-anisotropy models is well known, as it provides the
key step in the argument of Pelcovits, Pytte and Rudnick.\cite{PPR}

In the last few years, Monte Carlo calculations have revealed that for the
$n=2$ case (where $n$ is the number of spin components) there is a low
temperature phase with power-law decay of two-spin correlations (which will
be referred to as the QLRO phase) for both the random uniaxial
anisotropy\cite{Fi2} and random field\cite{Fi3} in three dimensions.  The
QLRO phase is characterized by the small-wavenumber behavior of the magnetic
structure factor
\begin{equation}
{\rm S}({\bf k}) = [ \langle |{\bf M} ({\bf k})|^2 \rangle ] \quad ,
\end{equation}
where the expectation value $\langle ~ \rangle$ indicates a thermal
average and [ ] is an average over the alloy disorder.  ${\bf M} ({\bf k})$
is the Fourier transform of the magnetization.  With the random field it was
found that in the QLRO phase the small {\bf k} behavior of S({\bf k}) goes
as ${\bf k}^{-3}$, and with the random uniaxial anisotropy it goes as
${\bf k}^{-2.4}$.  A spin-wave theory analysis would predict that this
behavior should be the same.  Thus the fact that the power law is found to
be different in the two cases demonstrates that vortex lines are controlling
the behavior for $n=2$ with random anisotropy.  The $n=2$ case with
three-fold (or higher) random anisotropy retains a stable FM phase in three
dimensions.\cite{Fi4}

Other papers which indicate the presence of QLRO for $n=2$ are the Monte
Carlo calculation of Gingras and Huse\cite{GH} for the random field case,
the high temperature susceptibility series of Fisch and Harris\cite{FH} for
the random anisotropy case, and the replica-symmetry breaking analysis of
Giamarchi and Le Doussal.\cite{GLD}  The latter work actually studies the
closely related problem of randomly pinned vortex lines in a dirty type-II
superconductor.\cite{Lar}

The existence of QLRO for $n=2$ is not a big surprise, given our previous
experience with the Kosterlitz-Thouless phase in the (nonrandom)
two-dimensional XY model.  However, the Aharony-Pytte-Goldschmidt\cite{AP,GA}
calculations find an infinite-susceptibility phase for any finite value of
$n$, independent of the existence of any topological defects.  Recently, a
Monte Carlo calculation by the current author\cite{Fi5} found a QLRO phase
with an S({\bf k}) which goes as ${\bf k}^{-3}$ for the $n=3$ model in a
random field in three dimensions.  Here we will extend this work to the case
of $n=3$ spins with random anisotropy, and we will find the same result for
the small {\bf k} behavior of S({\bf k}) as for the random field.

\section{DISCRETIZED HEISENBERG MODEL WITH RANDOM ANISOTROPY}

The Hamiltonian we study in this work is the obvious modification of the
one previously used for the $n=3$ random field model.\cite{Fi5}  It consists
of a FM Heisenberg exchange term, with the addition of a cubic single-ion
anisotropy term and a random-anisotropy term due to the alloy disorder.
Thus the form of the Hamiltonian is
\begin{equation}
 H = - J\sum_{\langle ij \rangle} \sum_{\alpha = 1}^3 S_{i}^\alpha
    S_{j}^\alpha - K\sum_i \sum_{\alpha = 1}^3 ( S_{i}^\alpha )^4
    - D\sum_{i^\prime} (({\bf S}_{i^\prime}
    \cdot {\bf n}_{i^\prime})^2 - 1 )\, ,
\end{equation}
where the sites $i$ form a simple cubic lattice and $\langle ij \rangle$
indicates a sum over nearest neighbors.  The $\alpha$ are spin indices, each
${\bf n}_{i^\prime}$ is an independently chosen random unit vector, and the
$i^\prime$ sites are a randomly chosen subset of the lattice containing a
fraction $x$ of the sites.

The $i^\prime$ sites represent the sites of the RE atoms, while the
remaining sites of the lattice contain the TM atoms.  This ignores the
fact that the RE-TM alloys of interest here are amorphous rather than
crystalline, but it is a reasonable first approximation.  Since the atoms
are assumed to be immobile, the ${\bf n}_{i^\prime}$ do not change with
time.  In this work we will study the limit in which $D$ is taken to
infinity.  The random anisotropy term is then a projection operator, and
the spin on each of the $i^\prime$ sites is restricted to two states,
parallel and antiparallel to the vector ${\bf n}_{i^\prime}$.  For
simplicity, we will assume that all of the spins are unit vectors and that
the exchange couplings $J$ between all pairs of nearest neighbor spins are
identical.  Our assumed values for the couplings are not close to the
actual values in RE-TM alloys,\cite{HAS} but they should still give the
qualitative behavior correctly, due to universality.

There are several reasons for including the cubic single-ion anisotropy term
in the Hamiltonian.  The first is a matter of convenience.  In order to
improve the efficiency of the Monte Carlo program,\cite{Fi5,Rap} and to make
it easy to store states of the system for later analysis, we are going to
discretize the phase space of the model.  For each spin variable ${\bf S}_i$
and each random anisotropy axis ${\bf n}_{i^\prime}$, we restrict the
allowed states to be the twelve [110] unit vectors.  We will refer to this
discretization as the ${\bf O}_{12}$ model.  This discretization induces an
effective cubic single-ion anisotropy term in the Hamiltonian.  The second
reason is that the experimental samples of amorphous RE-TM alloys are
sputtered films which have some growth-induced anisotropy.\cite{CK}
Understanding the effects of this weak anisotropy is a worthwhile exercise.
The existence of the growth-induced anisotropy is crucial for the use of
these films as magneto-optical memory devices.\cite{HH}  The third reason is
that besides the amorphous RE-TM alloys, we are also interested in
crystalline spin-glass alloys like CuMn,\cite{Fi2,Wer} where a cubic
single-ion anisotropy is present in the real materials.

Another discretized three-dimensional $n=3$ Hamiltonian which combines cubic
single-ion anisotropy and random uniaxial anisotropy has been studied
previously.\cite{Fi6}  In that case only six states (the [100] unit vectors)
were used.  Although the six-state discretization is too coarse to serve as
a quantitative approximation to the Heisenberg model, some indications of a
QLRO phase were found in that work.

A discussion from a renormalization-group viewpoint of the effects of having
both a crystalline anisotropy and random uniaxial anisotropy has been given
by Mukamel and Grinstein.\cite{MG}  They argued that a QLRO phase could
exist in this case between the FM and the paramagnet (PM), as was later
found by the Monte Carlo calculation for the $n=2$ case.  They also
discussed the difference between a discretized model in which all of the
allowed axes are mutually orthogonal, such as the six-state model for $n=3$,
and finer discretizations such as the ${\bf O}_{12}$ model, which are
expected to provide approximations to a model without the crystalline
anisotropy.

Since our ${\bf O}_{12}$ discretization of the spin variables automatically
builds a cubic anisotropy into the free energy, and we are primarily trying
to understand the qualitative aspects of the ordering and not attempting a
quantitative model of a particular experiment, we do not keep the $K$ term
in the Hamiltonian explicitly.  Thus, the Hamiltonian is reduced to the
simple form
\begin{equation}
 H = - J\sum_{\langle ij \rangle} {\bf S}_i \cdot {\bf S}_j \quad .
\end{equation}
The random anisotropy term has been reduced to the constraint\cite{JK} that
for each site in the $i^\prime$ set the spin ${\bf S}_{i^\prime}$ has only
two allowed states, either parallel or antiparallel to the vector
${\bf n}_{i^\prime}$.

\section{MONTE CARLO CALCULATION}

Because all of the ${\bf S}_i$ are chosen from the [110] states, Eq.~(3) has
the useful property that the energy of every state is an integral multiple
of 1/2.  Thus it becomes easy to write a Monte Carlo program to study
Eq.~(3) which uses integer arithmetic to calculate energies.  This, plus the
fact that that each spin has only twelve possible states, gives substantial
improvements in performance over working with the general form of Eq.~(2),
for both memory size and speed.  It is also be possible to use integer
arithmetic if $D$ is chosen to be an integer.\cite{Fi6}

The Monte Carlo program used two linear congruential pseudorandom number
generators.  In order to avoid unwanted correlations, the decision of
whether or not a state was in the $i^\prime$ set was done using one of the
generators, and the choice of the vector ${\bf n}_{i^\prime}$ was made
using the other.  A heat bath method was used for flipping the spins, which
at each step reassigned the value of a spin to one of its two allowed states
if it was a member of the $i^\prime$ set, or to one of the twelve [110]
states if it was not, weighted according to the Boltzmann factors and
independent of the prior state of the spin.

$L \times L \times L$ simple cubic lattices with periodic boundary
conditions were used throughout.  The values of $L$ used ranged from 16 to
64.  Away from any $T_c$ the samples were typically run for 10,240 Monte
Carlo steps per spin (MCS) at each $T$, with sampling after each 10 MCS.
Near a $T_c$ they were run several times longer.  The initial part of each
data set was discarded for equilibration.  In some cases, two different
random field configurations with a given $L$ were studied for a given $x>0$.
This gives a rather crude estimate of the $L$ dependence of the various
thermodynamic properties.  We are forced to go to large $L$ for this problem
by crossover effects.  To do a high precision finite-size-scaling analysis
would require studying many samples at each $L$, which is very
time-consuming for large $L$. 

Due to our discretization of the spins, in a ground state with $x \leq 0.25$
essentially all of the spins are as aligned as possible along one of the
[110] directions, consistent with the restriction that each $i^\prime$ spin
be in one of its two allowed states.  Thus, in these cases it is easy to
equilibrate the system at low temperatures by starting from an ordered
configuration.  For $x=0.25$ a sample with small $L$ will spontaneously
nucleate a [110] FM state upon cooling, but for $L=64$ this does not happen
in the time available.  When an $L=64$ sample with $x=0.25$ is slowly cooled
to $T/J=0.5625$ it nucleates a polydomain state.  For $x \geq 0.5$, where
the ground state is not a [110] FM state, it becomes extremely difficult to
equilibrate large lattices at low temperatures.

In the absence of any external field, the rotation of $\bf M$ between
different [110] directions is a slow process.  In the presence of the random
anisotropy the different [110] ferromagnetic Gibbs states have different
energies.  Because all of these twelve minima are equivalent, on the
average, there is no need for the Monte Carlo program to average over them.
If the system is started in a high-energy [110] direction, it will
eventually jump to a more favorable direction, unless $T$ is so low that
this does not happen in the time available.

\section{NUMERICAL RESULTS}

For $x=0$ there is no random anisotropy, and results for this case were
presented earlier.\cite{Fi5}  Monte Carlo simulations with the random
anisotropy were obtained for $x=0.125$, 0.25, 0.5, 0.75 and 1.0.  A
approximate picture of the phase diagram obtained from these results is
shown in Fig.~1.  The limit of stability of the [110] FM ground state is
between $x=0.25$ and $x=0.5$.  The QLRO phase exists for $x \leq 0.5$, but
is unstable at $x=0.75$.

The [111] FM phase, which is stable at $x=0$,\cite{Fi5} should also extend
to small positive values of $x$.  The domain walls in this phase are broad
and have a low cost in free energy.  It is difficult to obtain meaningful
numerical results for very small $x$, due to crossover effects.\cite{BD}
The [111] FM-QLRO phase boundary was not observed directly, and its
existence is shown in Fig.~1 as a dotted line.  Although for small $x$ the
effective cubic anisotropy induced by our discretization favors the [111]
directions, for $x$ near 0.5 the effective anisotropy favors the [100]
directions.  Consequently, near $x=0.25$ the effective cubic anisotropy is
very small.  There is probably a [100] FM low temperature phase near
$x=0.5$.  However, because it is so difficult to equilibrate large $L$
samples in this region of the phase diagram, the details are very uncertain.

The QLRO-to-PM transition is second order for $x \leq 0.5$.  The energy at
$T_c$ remains nearly constant along this part of the transition line,
decreasing very slowly as $x$ increases.  The shift of $T_c$ with $x$ is
surprisingly small here; at $x=0$
\begin{equation}
 {d \over dx}\biggl({{T_c (x)} \over {T_c (0)}}\biggr)=-0.12\pm 0.02 \quad .
\end{equation}
Similar effects were observed in the $n=2$ random anisotropy case,\cite{Fi2}
but the random field cases are significantly different.\cite{Fi5}  At some
$x > 0.5$ $T_c$ begins to drop rapidly, and by $x=0.75$ the QLRO phase has
disappeared.  The details of the sharp drop in $T_c$ are not clear, and they
may depend on the presence of the effective cubic anisotropy, without which
the [100] FM phase would be stable.

Because in the presence of the random anisotropy the expectation values of
the energies of different bonds are not the same, one certainly does not
expect the energy at $T_c$ to remain exactly constant, independent of $x$.
Unlike the random field, however, which causes spins to have nonvanishing
expectation values even at high temperatures, the random uniaxial anisotropy
does not produce major qualitative changes in the PM phase.  Therefore, it
is not unreasonable that the limit of stability of the PM phase occurs at
about the same value of the nearest neighbor two-spin correlation function,
which is the energy in this model.  Recall that in a tree-diagram
summation the energy at $T_c$ of the classical $n$-vector ferromagnet
depends only on the lattice structure, and is independent of the number of
spin components.

For the $n=2$ random anisotropy model it was found\cite{Fi2} that the QLRO
phase exists on a simple cubic lattice\cite{note1} even for $x=1$, but here
we see that QLRO is less stable for $n=3$ random anisotropy.  Although the
$n=4$ case is not (to the author's knowledge) of any experimental relevance,
it would be quite interesting to know if it is possible to have a QLRO phase
for small nonzero $x$ in that case, and for larger finite values of $n$.
The author sees no reason why this should not be so.  Some authors\cite{LD}
have argued for the importance of hedgehog excitations in three-dimensional
$n=3$ models, but this does not seem to hold up under detailed
scrutiny.\cite{HJ}  In a lattice model one can obtain equivalent effects by
replacing the hedgehog fugacity with a nontopological multispin interaction,
and this multispin interaction can be extended in a straightforward way to
the $n=4$ case.\cite{Fi7}

The behavior of the specific heat as $x$ is increased is shown in Fig.~2.
The data displayed were obtained by numerically differentiating the
calculated values of the energy with respect to $T$.  The specific heat was
also computed by calculating the fluctuations in the energy at fixed
temperature, yielding similar but noisier results.  We see that the data
for different samples with the same value of $x$ agree fairly well, although
some differences are visible near the phase transitions.

The sharp peaks which occur\cite{Fi5} for $x=0$ become rounded as $x$
increases, and they move to lower $T$.  At $x=0.125$ the behavior of the
specific heat at the QLRO-to-PM transition can be approximated by an
effective critical specific heat exponent $\alpha_{eff}= -0.45$, with an
amplitude ratio of 2.5.  The transition out of the [110] FM phase still
appears to be first order at $x=0.125$, although the latent heat is too
small to measure accurately.\cite{note2}  It is likely that this transition
actually goes into the [111] FM phase.  At $x=0.25$ there is substantial
hysteresis at the [110] FM-to-QLRO transition, which makes an accurate
determination of the equilibrium thermodynamic properties near this
transition impossible.  The specific heat near the QLRO-to-PM transition at
$x=0.25$, shown in Fig.~2(b), is remarkably similar to the specific heat
recently reported for real amorphous RE-TM films.\cite{HAS}  The effective
value of the critical specific heat exponent is now $\alpha_{eff}= -0.6$,
and the amplitude ratio is again about 2.5.  It is well known\cite{RMSP}
that the introduction of randomness gives rise to effective critical
exponents that appear to vary with $x$.  By $x=0.5$ the specific heat
peak has become rather smeared out, but it still seems to be centered at
the QLRO-to-PM transition.  The shoulder on the low temperature side of the
peak in Fig.~2(c) may be due to a transition from the QLRO phase into a
[100] FM phase.

Looking at the dependence of $\langle |{\bf M}| \rangle$ on $x$ and $L$
provides additional insight.  The data for $x=0.25$ and $x=0.5$ are
shown in Fig.~3.  In the [110] FM phase, $\langle |{\bf M}| \rangle$ is
almost independent of $L$, except very close to $T_c$.  In the [111] and
[100] FM phases, $\langle |{\bf M}| \rangle$ becomes independent of $L$ only
when $L$ is larger than the domain-wall thickness.  In practice we do not
satisfy this condition, and for accessible $L$ it becomes very difficult to
distinguish the [111] and [100] FM phases from the QLRO phase.  In the QLRO
phase, $\langle |{\bf M}| \rangle$ decreases slowly as $L$ increases,
probably decaying as $1/ \log (L)$.  In the PM phase,
$\langle |{\bf M}| \rangle$ decreases as $(L/\xi)^{-3/2}$, where $\xi$ is
the ferromagnetic correlation length.

In Fig.~3 we see that for both values of $x$ the variation of
$\langle |{\bf M}| \rangle$ with $T$ becomes increasingly sharp as $L$
increases.  If one looks at finite-size scaling plots (not shown), there is
substantial scatter due to sample-to-sample fluctuations.  However, in both
cases the finite-size scaling is consistent with a divergence of $\xi$ as
$T$ approaches $T_c$ in the PM phase, with an effective value of $\beta/\nu$
(and therefore of $\eta$) which is indistinguishable from the standard $n=3$
Heisenberg critical point value.\cite{LGZJ}  The data can also be fit with
$\eta = 0$.  The effective values of $\nu$ are, however, about 0.8 at
$x=0.25$ and 1.0 at $x=0.5$.  Note that if one uses the effective value of
$\nu$ found here for $x=0.25$ and the effective value of $\alpha$ found in
the specific heat, the Josephson relation $d \nu = 2 - \alpha$ is satisfied
within the accuracy of the estimates.  There is no reason, however, why
effective exponents should satisfy scaling relations exactly.

We can get valuable information by looking at the magnetic structure factor
S$({\bf k})$.  The structure factor can be measured by X-ray and neutron
scattering experiments.  Near $T_c$ the small-wavenumber behavior of the
structure factor is expected to have the form
\begin{equation}
{\rm S}({\bf k}) \approx A/(1/\xi^2 + |{\bf k}|^2)^{1-\eta /2} \quad .
\end{equation}
As $T$ approaches $T_c$ in the PM phase, $\xi$ is expected to diverge like
$( T - T_c )^{-\nu}$.  S$({\bf k})$ at $T_c$ for an $L=64$ sample with
$x=0.25$ is shown on a log-log plot in Fig.~4.  The exponent $\eta$ is seen
to be very close to zero; this is the same result that was found for this
exponent in the $n=2$ random anisotropy case.\cite{Fi2}

Inside the QLRO phase S$({\bf k})$ appears to take on the form
\begin{equation}
{\rm S}({\bf k}) \approx A/(1/\xi^2+|{\bf k}|^{2}) +
B/(1/\xi^2+|{\bf k}|^{2})^{1-\eta_0 /2} \quad ,
\end{equation}
with $\eta_0=-1$, independent of $T$.  The $B$ term has replaced the
$\delta$-function which would be found in the structure factor of a
ferromagnet.  The coefficient $B$ goes to zero continuously as $T$
approaches $T_c$ from below, presumably with an exponent $2 \beta$.  At
$x=0.25$ where the effective cubic anisotropy is small, $\xi$ is found to be
immeasurably large in the QLRO phase, which means that it must be large
compared to $L=64$.

This is shown in Fig.~5, which displays S({\bf k}) data from the same $L=64$
sample as in Fig.~4, but for lower $T$.  The data set shown with the +
symbols was obtained by slowly cooling the lattice from above $T_c$ to
$T/J=0.6875$.  After discarding the initial part of the run, this state
appears stationary on a timescale of $50,000$ MCS.  It has
$\langle |{\bf M}| \rangle = 0.419$, and an energy of $-1.9992$.  The fit of
this data to a straight line with a slope of $-3$ is excellent.  The
remaining data shown here were generated using a cold start initial
condition, with the direction of {\bf M} chosen to be approximately the same
as for the slowly cooled state.  At $T/J=0.5$, which is within the [110] FM
phase, $\langle |{\bf M}| \rangle = 0.812$ and S({\bf k}) is rather flat at
nonzero {\bf k}.  Upon warming this state up to $T/J=0.5625$, which is
approximately equal to the FM-QLRO transition point,
$\langle |{\bf M}| \rangle$ has decreased to 0.757, and S({\bf k}) has
increased at nonzero {\bf k}.

Using a cold start initial condition at $T/J=0.6875$ results in a state
which does not relax to equilibrium on accessible timescales.  The data
shown in Fig.~5 with the diamond symbols fall on top of the equilibrium
(slow-cooled) data for ${\bf k} \geq 0.1$, but they are flat at smaller
{\bf k}, where relaxation times are long.  These data have
$\langle |{\bf M}| \rangle = 0.599$ and an energy of $-2.0045$, but there is
a clear trend of decreasing $\langle |{\bf M}| \rangle$ and increasing
energy with time.  The rounded peak\cite{Wer} and slow relaxation\cite{NSLS}
of this state are quite reminiscent of the field-cooled state in the
spin-glass phase of CuMn alloys.

For $x \geq 0.75$ the ferromagnetic correlation length does not become
larger than accessible values of $L$ at any temperature, and S({\bf k}) can
be fit by a simple Lorentzian form ({\it i.e.} setting $\eta =0$ in
Eq.~(5)).  This is shown in Fig.~6.  For $x=0.75$ we display data taken at
$T/J=1.15625$ for an $L=64$ sample, using both slow cooling and cold start
initial conditions.  This value of $T$ is slightly below what our
extrapolation from small $x$ would lead us to expect $T_c$ would be.
Although there is a substantial peak in S({\bf k}), the value of $\xi$
(measured in lattice units) appears to be $25 \pm 5$, and it does not
increase as $T$ is lowered.

Data for two $L=32$ samples with $x=1$ at $T/J=0.75$ using slow cooling are
shown in Fig.~6(b).  Here the value of $\xi$ is $11 \pm 1$, in excellent
agreement with earlier estimates for the $x=1$ model with $n=3$ isotropic
random anisotropy.\cite{JK,Fi1}  For $x=1$, $\xi$ becomes essentially
temperature-independent below about $T/J=0.9$, but there is no evidence of
singular behavior at any $T$.  Given the recent results of Migliorini and
Berker,\cite{MB} the author does not believe that there is any phase
transition at $T>0$ in this model when $x \geq 0.75$.  Although no attempt
was made to find the exact ground states of these samples, the estimated
ground state energy for the ${\bf O}_{12}$ model at $x=1$ is $-1.12 \pm 0.01
J$, essentially equal to its value in the isotropic random anisotropy case.

\section{DISCUSSION}

The integral of S({\bf k}) over {\bf k} is proportional to the total
neutron-scattering cross-section, which is finite.  Because $|{\bf k}|^{-3}$
is not an integrable singularity in three dimensions, either $\xi$ must
become finite, albeit extremely large, below $T_c$ where $B$ is nonzero, or
else $\eta_0$ must be modified slightly to make the integral converge.  In
the presence of some crystalline or growth-induced nonrandom anisotropy, it
is natural to expect that $\xi$ becomes finite, just as it does in a
nonrandom ferromagnet below $T_c$.\cite{Fi7}  Although it has become
traditional to fit neutron-scattering data for S({\bf k}) in random-field
and random-anisotropy magnets to a Lorentzian plus Lorentzian$^2$ form, this
is based on a theoretical preconception.  It has been known for a long time
that Eq.~(6) will serve to fit the data in some cases,\cite{Yetal} while in
others\cite{PRA,Yetal} the 3 in the exponent should be replaced by 2.4.
A value of 2.4 in this exponent is precisely the value found for the $n=2$
random-anisotropy model,\cite{Fi2} and thus should be expected in the
presence of easy-plane anisotropy.  The $|{\bf k}|^{-3}$ behavior of
S({\bf k}) for $n=3$ is the same exponent for both the random-anisotropy and
random-field cases.  Therefore, we have no reason to believe that the
presence of hedgehog singularities is essential for the QLRO when $n=3$.
There seems to be no reason why QLRO should not exist for larger finite
values of $n$, as originally predicted by Aharony and Pytte.\cite{AP}

The magnetic susceptibility is
\begin{equation}
\chi^{\alpha \beta} = (NT)^{-1} \sum_{i, j = 1}^N
       [ \langle S_{i}^\alpha S_{j}^\beta \rangle ] -
       [ \langle S_{i}^\alpha \rangle \langle S_{j}^\beta \rangle ] \quad .
\end{equation}
The Aharony-Pytte-Goldschmidt\cite{AP,GA} analysis predicts that at small
wavenumbers its Fourier transform, $\chi ({\bf k})$ will behave like
$|{\bf k}|^{-2}$.  If time-reversal symmetry is unbroken, then the
$[ \langle S_{i}^\alpha \rangle \langle S_{j}^\beta \rangle ]$ terms all
vanish, and the trace of $\chi ({\bf k})$ is proportional to S({\bf k}).
Thus Aharony-Pytte-Goldschmidt implicitly predicts that S({\bf k}) goes like
$|{\bf k}|^{-2}$ in the QLRO phase, which is not correct.  Its prediction
for $\chi ({\bf k})$ in the QLRO phase is, however, probably correct in the
absence of any crystalline anisotropy.  To check this numerically would
require very long runs, in order to compute the
$[ \langle S_{i}^\alpha \rangle \langle S_{j}^\beta \rangle ]$ terms
accurately.  This has not been attempted here.

It is likely that the reason why Aharony-Pytte-Goldschmidt fails to predict
the QLRO for the random-field case is also its incorrect handling of the
$[ \langle S_{i}^\alpha \rangle \langle S_{j}^\beta \rangle ]$ terms.  For
the random-field case these terms are nonzero even in the PM phase.

The FM phases shown in Fig.~1 exist because of the anisotropy induced by our
discretization.  It would be very helpful to repeat this calculation using
an alternative discretization, such as Rapaport's 30-state model,\cite{Rap}
which has icosahedral symmetry.  It would also be desirable to study a
continuous-spin binary alloy model with a fully isotropic distribution of
the random anisotropy.  This would be very difficult to manage for large
lattices, unless a way can be found to adapt the cluster Monte Carlo
algorithm\cite{Wol} to include the random anisotropy.  When $x>0$ the
Hamiltonian no longer has any planes of reflection symmetry, even though it
is still inversion-symmetric.

\section{CONCLUSION}

In this work we have used Monte Carlo simulations to study the ${\bf O}_{12}$
version of the diluted random-anisotropy ferromagnet in three dimensions.
We have found that there are two types of ordered phases, just as in the
$n=2$ case.  In addition to the anisotropy-stabilized ferromagnet, we find
an intermediate QLRO phase displaying a $|{\bf k}|^{-3}$ behavior of the
magnetic structure factor.  This is the same behavior which has been found
for $n=3$ random-field model.  The critical exponent $\eta$, which
characterizes the two-spin correlations on the QLRO-to-PM critical line,
has a value which is indistinguishable from zero.  The results should be
applicable to a variety of experimental systems, including amorphous RE-TM
alloys and CuMn-type spin-glass alloys.

\acknowledgments

This work was undertaken after the author was shown a preliminary version
of the results of Hellman, Abarra and Shapiro.  The author is greatful to
Frances Hellman for extensive discussions of these results.  He also thanks
Phil Anderson, Nihat Berker, Yadin Goldschmidt, Brooks Harris, David Huse
and Tom Lubensky for helpful discussions.

\begin{figure}
\caption{
Phase diagram of the dilute random-anisotropy ${\bf O}_{12}$ model on simple
cubic lattices, showing the paramagnetic (PM), ferromagnetic (FM), and
quasi-long-range order (QLRO) phases.  The plotting symbols show estimates
obtained from the Monte Carlo data.  The solid lines indicate first-order
transitions, and the dashed lines indicate second-order transitions.  The
dotted lines represent transitions which are inferred indirectly, and whose
locations are rather uncertain.}
\label{fig1}
\end{figure}

\begin{figure}
\caption{
Specific heat vs. temperature for the dilute random-anisotropy ${\bf O}_{12}$
model on $L \times L \times L$ simple cubic lattices.
(a)~$x=0.125$; (b)~$x=0.25$; (c)~$x=0.5$.}
\label{fig2}
\end{figure}

\begin{figure}
\caption{
Magnetization vs. temperature for the dilute random-anisotropy ${\bf O}_{12}$
model on $L \times L \times L$ simple cubic lattices.  The $y$-axis is
scaled logarithmically. (a)~$x=0.25$; (b)~$x=0.5$.}
\label{fig3}
\end{figure}

\begin{figure}
\caption{
Angle-averaged magnetic structure factor at the PM-to-QLRO transition for
the dilute random-anisotropy ${\bf O}_{12}$ model on a
$64 \times 64 \times 64$ simple cubic lattice with $x=0.25$, log-log plot.
The points show averaged data from 4 states sampled at 10,240 MCS intervals.
The line has a slope of $-2.00$.}
\label{fig4}
\end{figure}

\begin{figure}
\caption{
Angle-averaged magnetic structure factor near the [110]~FM-to-QLRO
transition for the dilute random-anisotropy ${\bf O}_{12}$ model on a
$64 \times 64 \times 64$ simple cubic lattice with $x=0.25$, log-log plot.
The data set shown with + symbols was obtained from a series of 4 states
which were obtained after cooling slowly from high temperature.  The other
data sets were obtained with a cold start initial condition.  The solid line
has a slope of $-3.00$, and the dashed line has a slope of $-2.00$.}
\label{fig5}
\end{figure}

\begin{figure}
\caption{
Angle-averaged magnetic structure factor at large $x$ and low $T$ for the
dilute random-anisotropy ${\bf O}_{12}$ model on $L \times L \times L$
simple cubic lattices, log-log plot.  The lines have a slope of $-2.00$.
(a)~$x=0.75$; (b)~$x=1.0$.}
\label{fig6}
\end{figure}

\begin{figure}
\epsfxsize=175pt 
\epsfbox{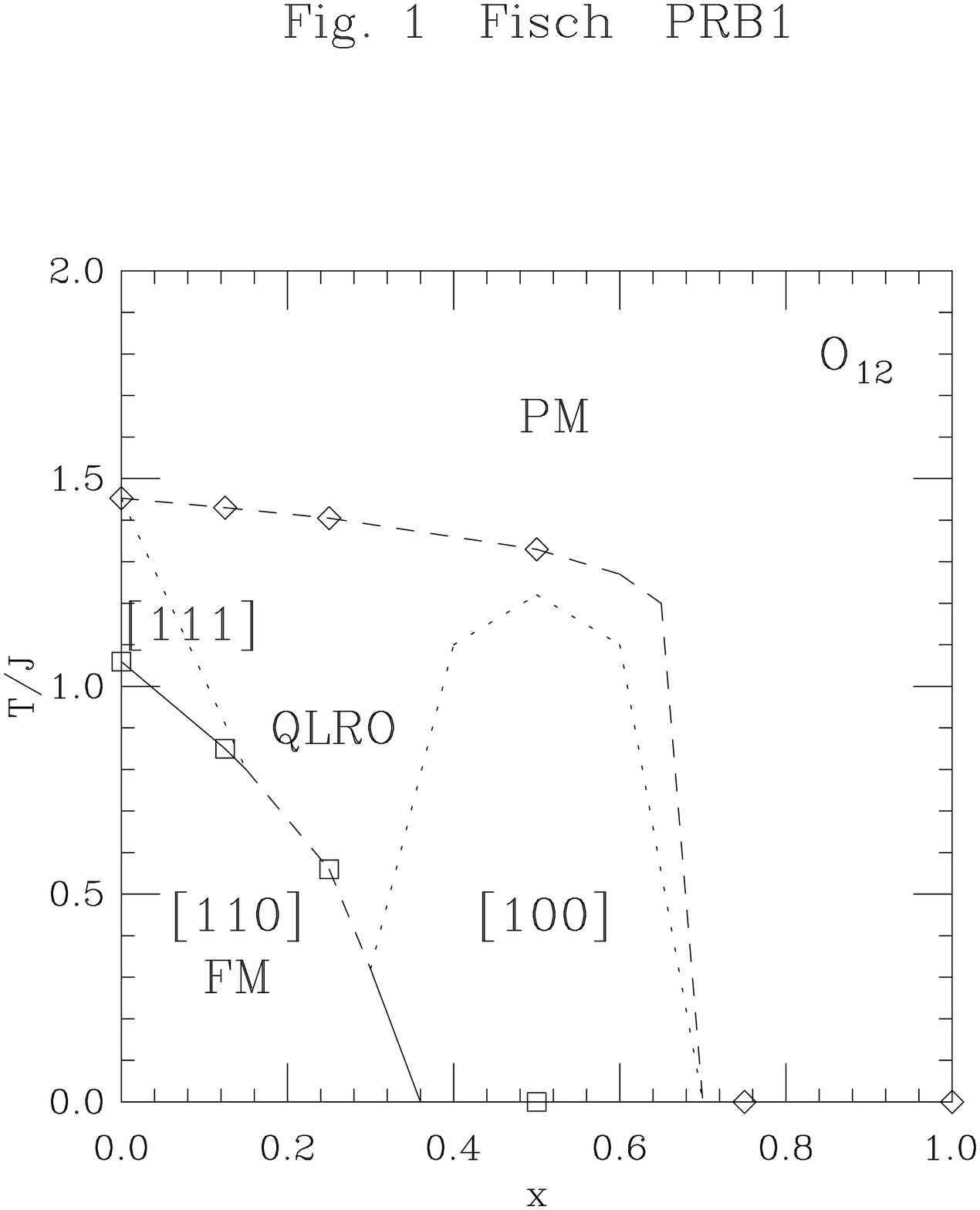}
\vskip 8truecm
\end{figure}

\begin{figure}
\epsfxsize=175pt 
\epsfbox{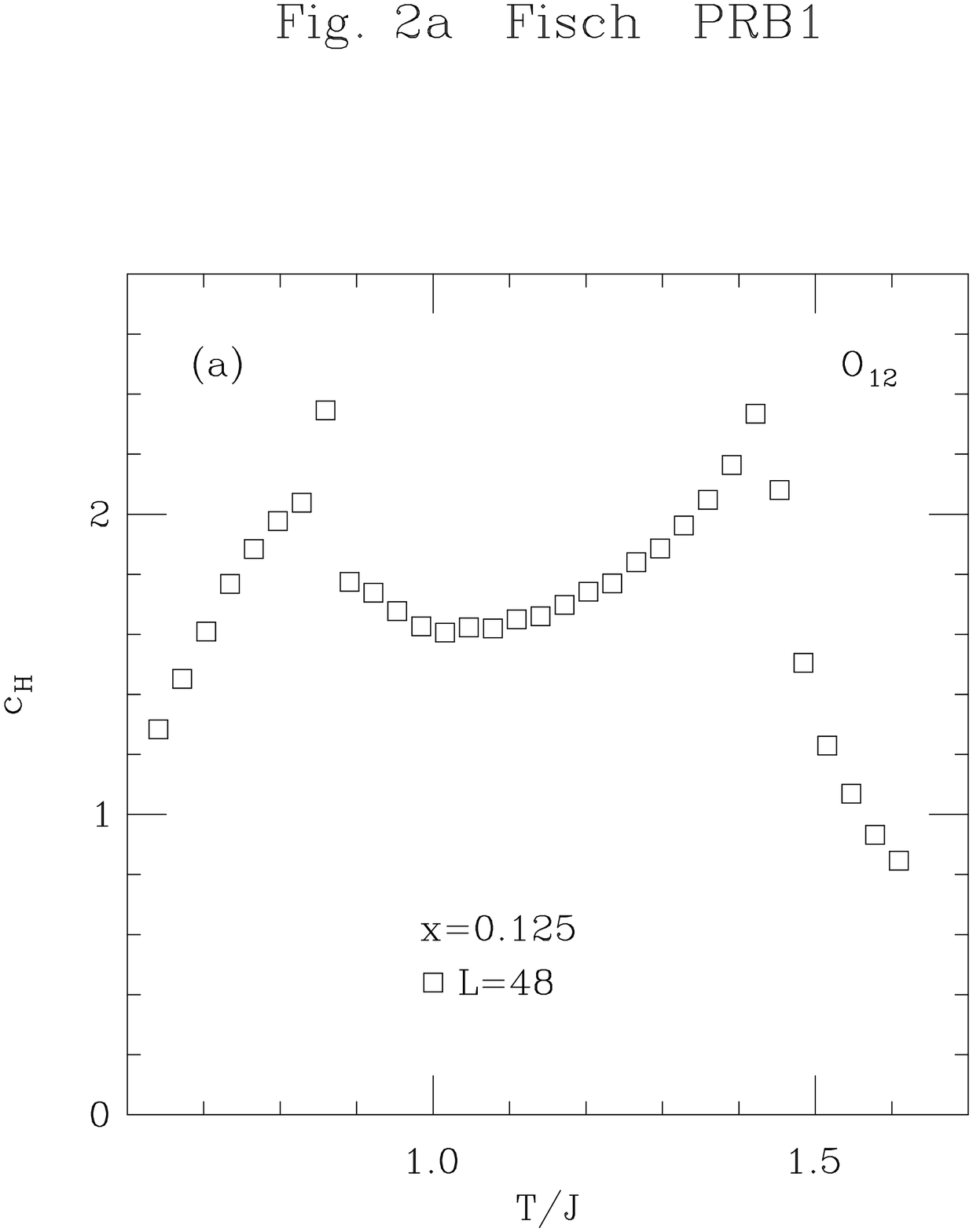}
\vskip 8truecm
\end{figure}

\begin{figure}
\epsfxsize=175pt 
\epsfbox{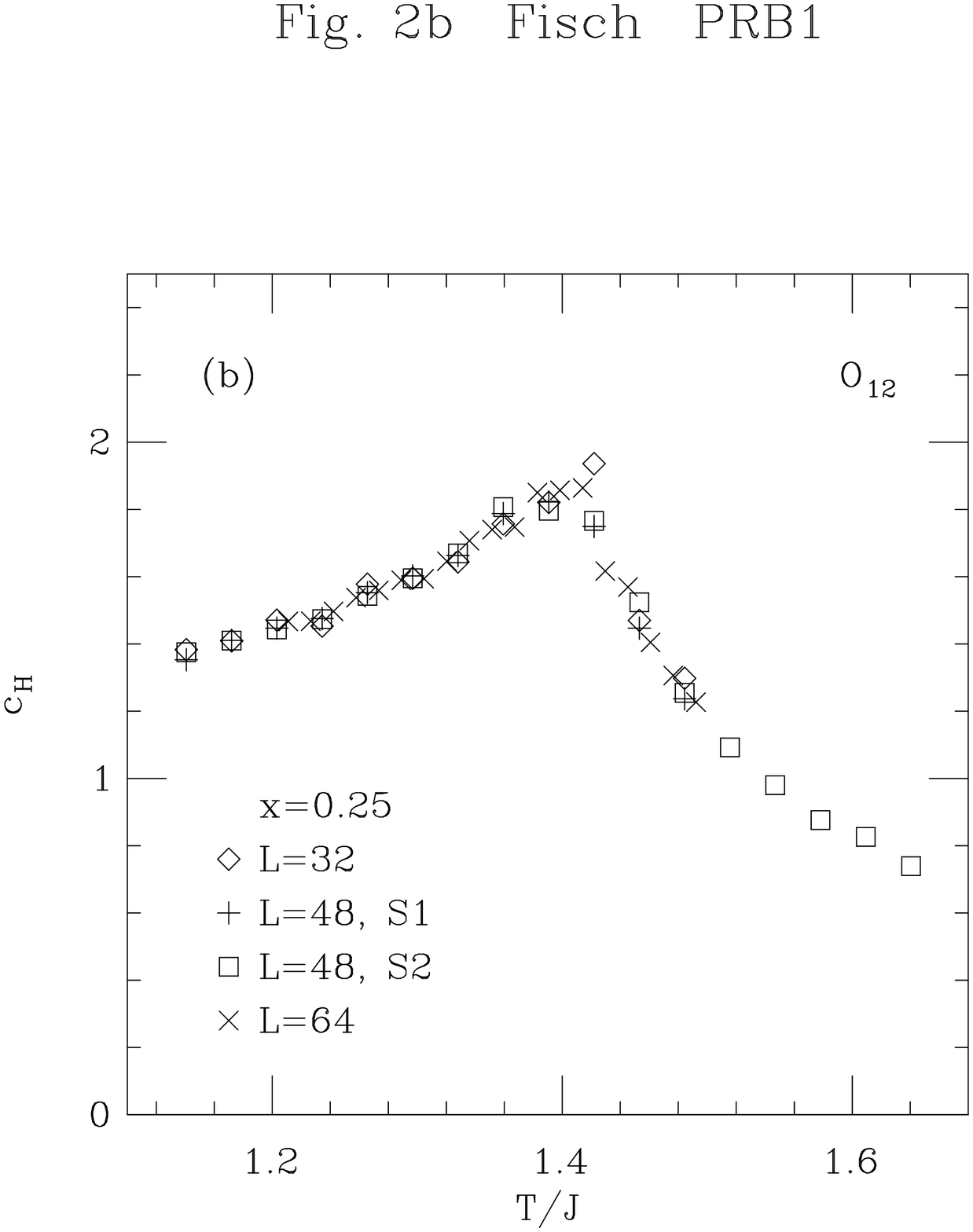}
\vskip 8truecm
\end{figure}

\begin{figure}
\epsfxsize=175pt 
\epsfbox{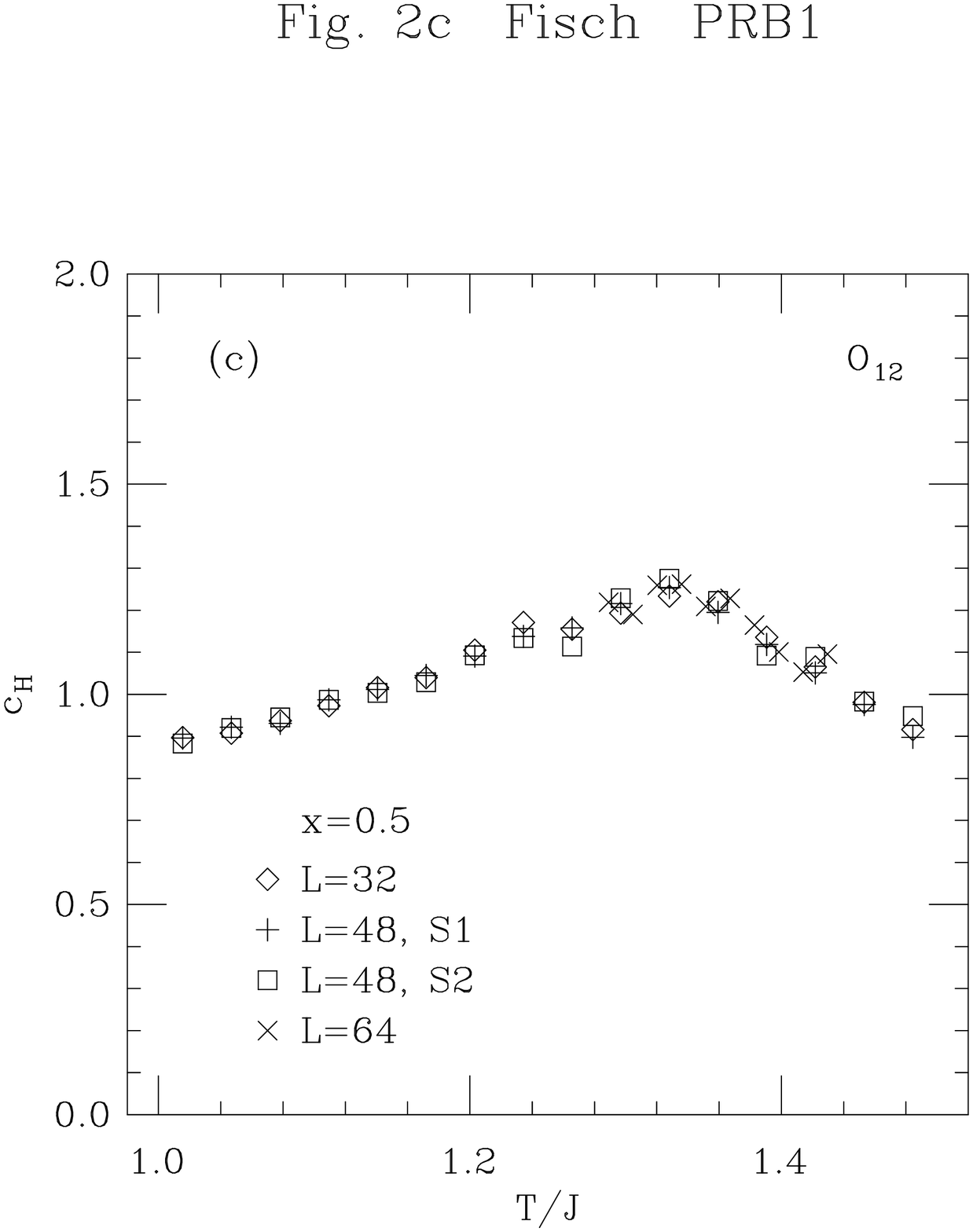}
\vskip 8truecm
\end{figure}

\begin{figure}
\epsfxsize=175pt 
\epsfbox{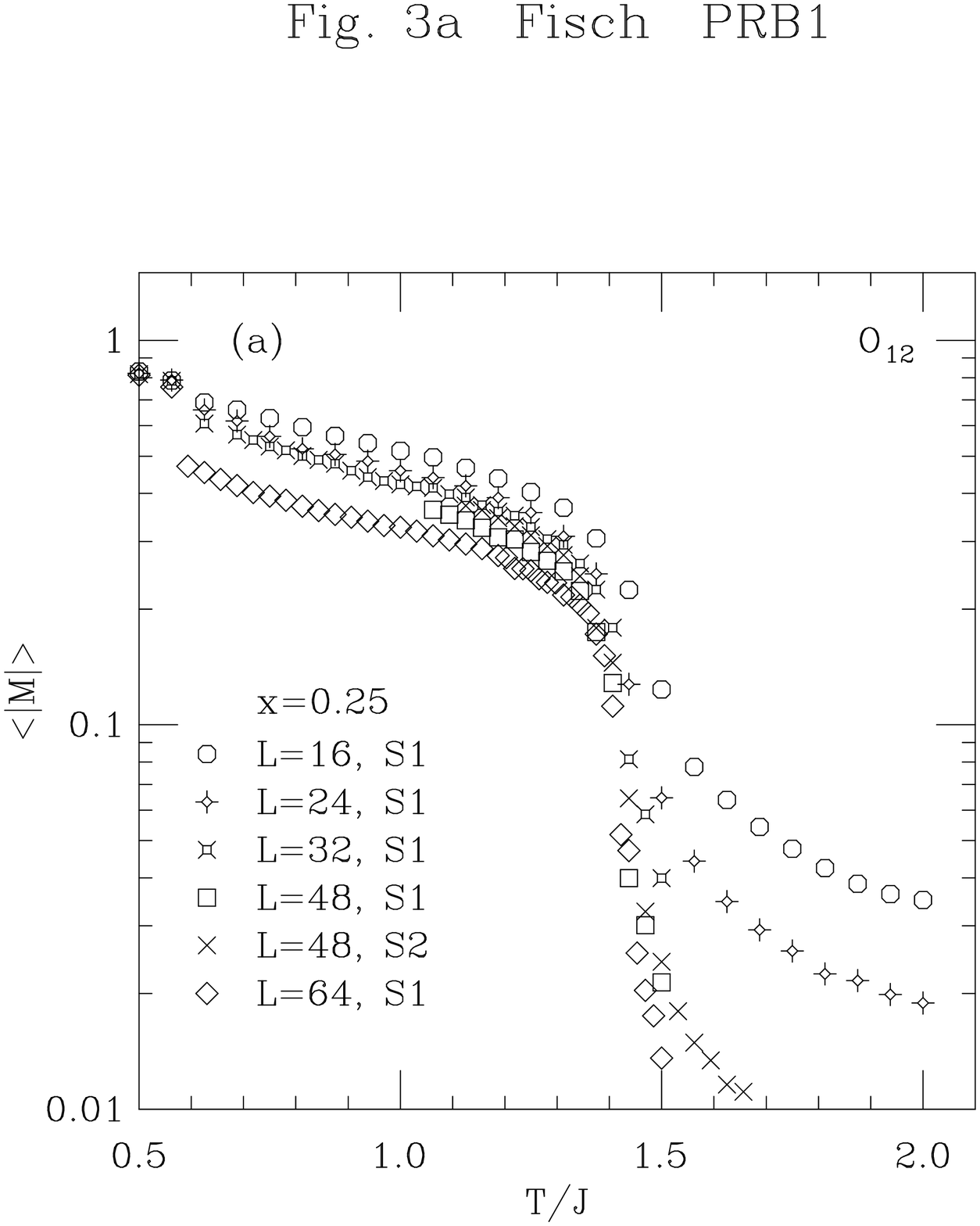}
\vskip 8truecm
\end{figure}

\begin{figure}
\epsfxsize=175pt 
\epsfbox{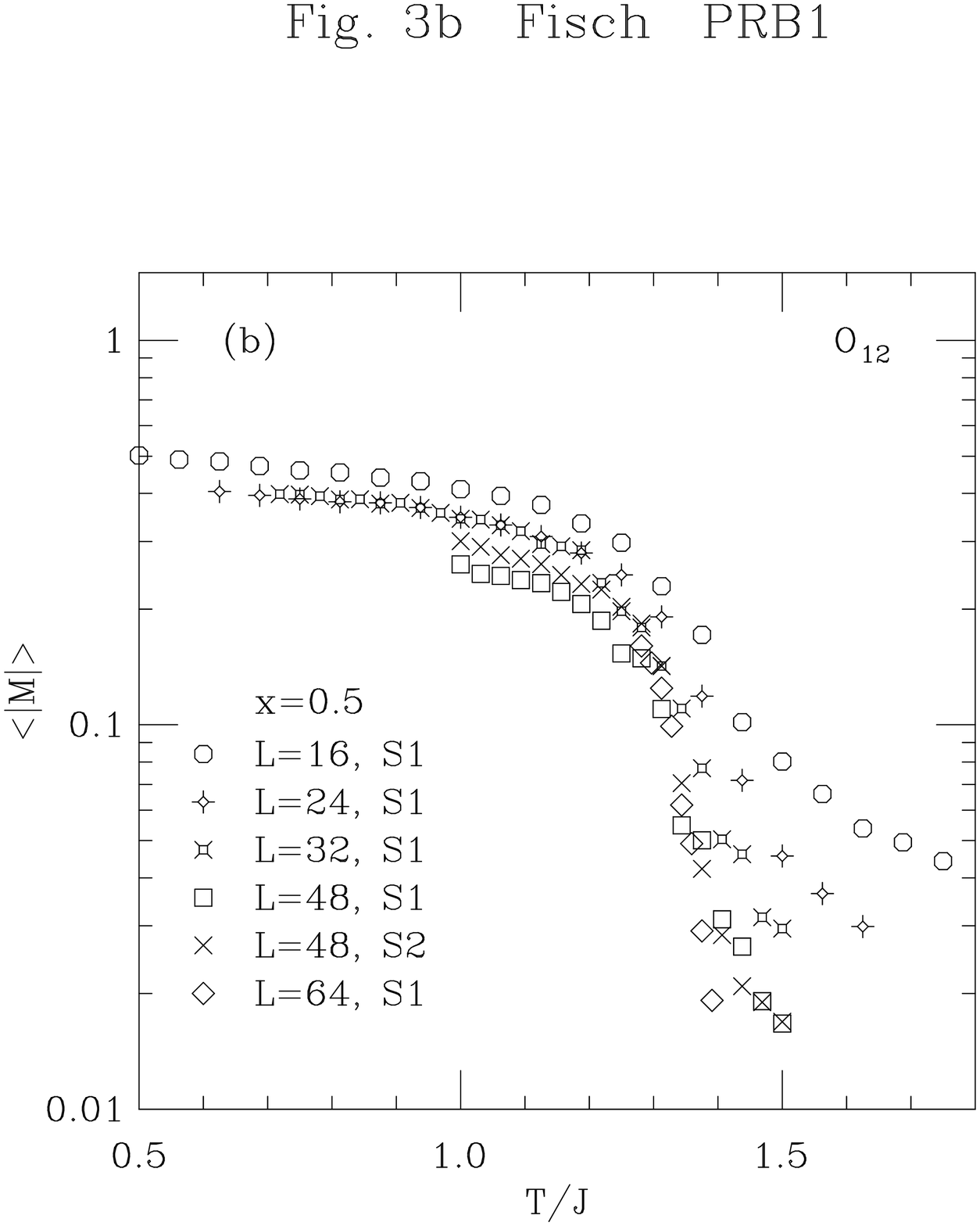}
\vskip 8truecm
\end{figure}

\begin{figure}
\epsfxsize=175pt 
\epsfbox{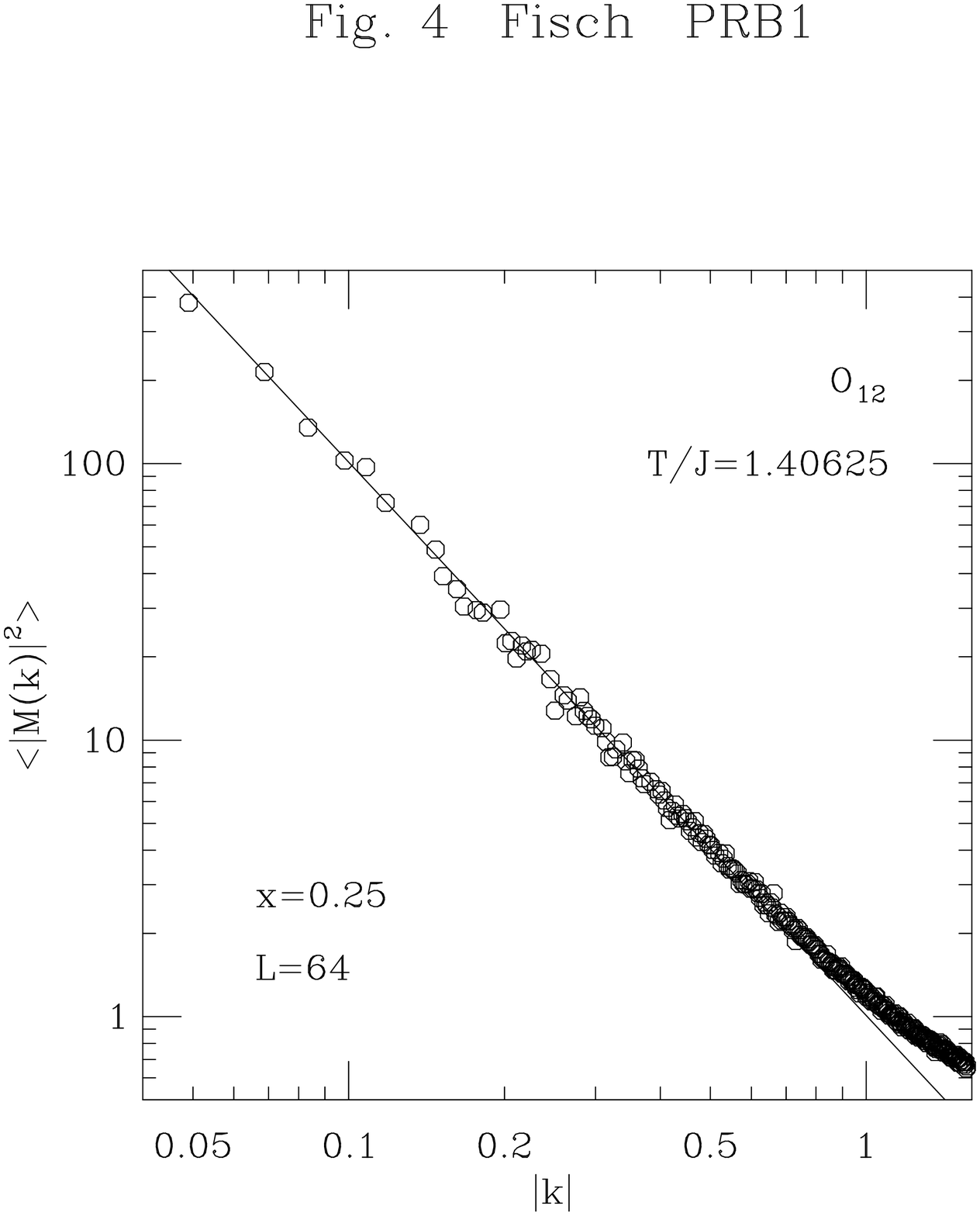}
\vskip 8truecm
\end{figure}

\begin{figure}
\epsfxsize=175pt 
\epsfbox{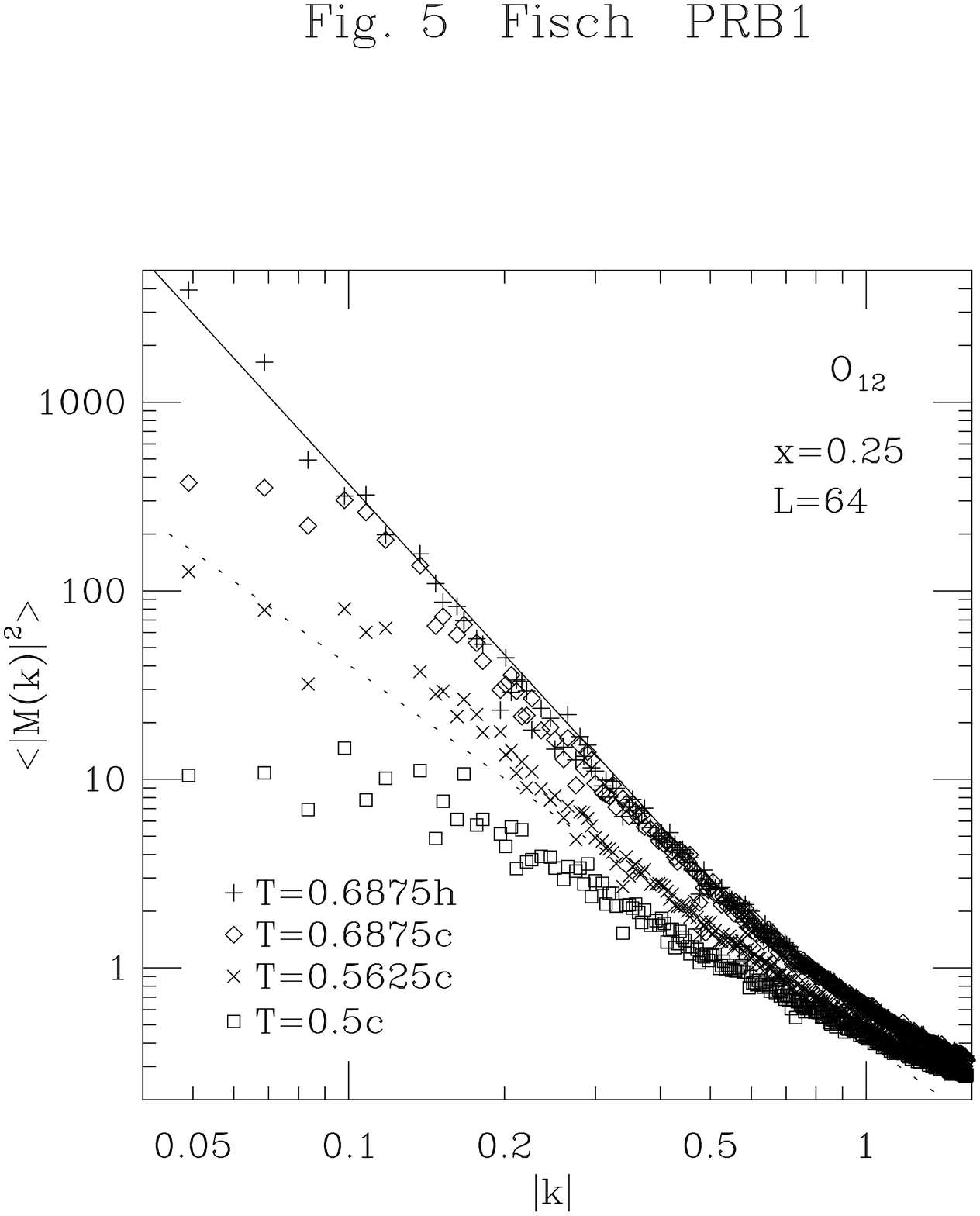}
\vskip 8truecm
\end{figure}

\begin{figure}
\epsfxsize=175pt 
\epsfbox{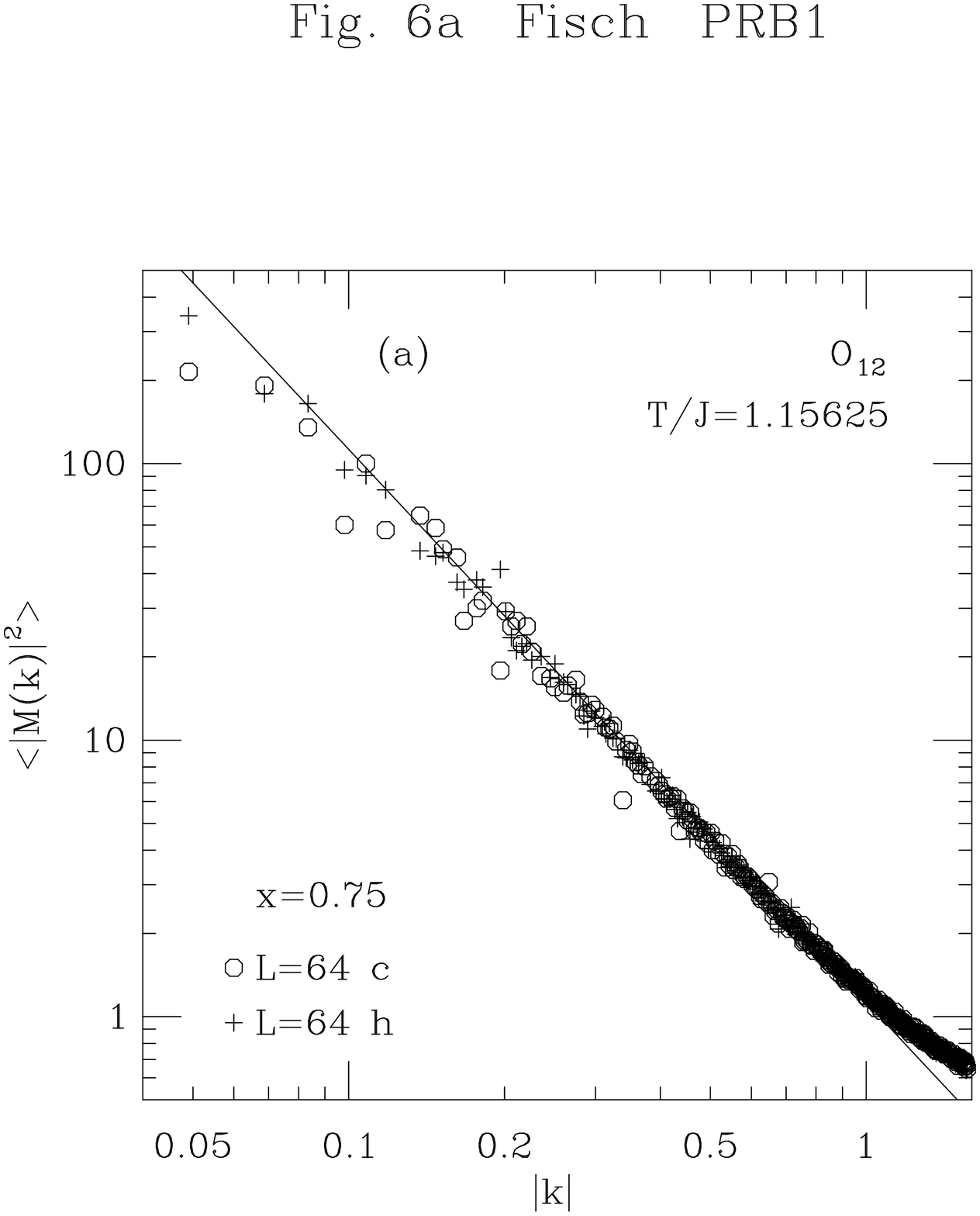}
\vskip 8truecm
\end{figure}

\begin{figure}
\epsfxsize=175pt 
\epsfbox{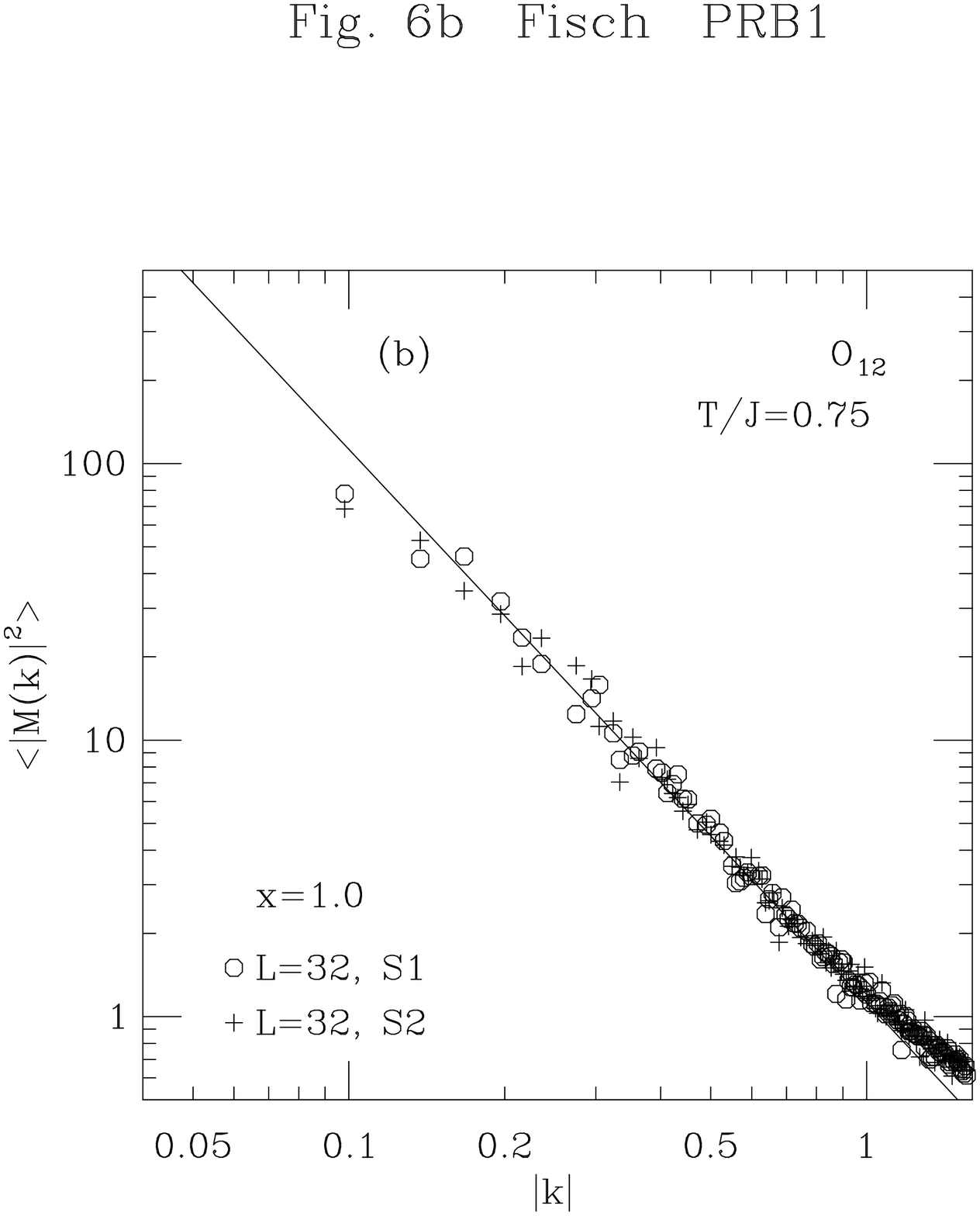}
\vskip 8truecm
\end{figure}


\begin{references}
\bibitem{note1}  
The existence of QLRO at $x=1$ is lattice dependent.  For example, if a
ferromagnetic second neighbor exchange coupling was added, then one might
expect a QLRO phase for $n=3$ at $x=1$.
\bibitem{note2}  
For the lattices sizes studied here, the sample-to-sample fluctuations are
comparable to the estimated latent heat itself.
\end{references}
\end{document}